\documentclass[prl,twocolumn,superscriptaddress,showpacs]{revtex4-1}
\usepackage{amsmath,amssymb}
\usepackage{bm}
\usepackage{hyperref}
\usepackage{subfigure}
\usepackage{graphics}
\usepackage{graphicx}
\usepackage{color}
\usepackage{soul}

\newcommand{\ket}[1]{\left| #1 \right \rangle}
\newcommand{\bra}[1]{\left \langle #1 \right|}

\hypersetup{
	pdfauthor={Radu S. Chicireanu},
	pdftitle={Differential Light Shift},
	colorlinks=true,
	linkcolor=blue,
	anchorcolor=blue,
	citecolor=blue,
	filecolor=blue,
	pagecolor=blue,
	urlcolor=blue
}

\begin{document}

\title{Differential Light Shift Cancellation in a Magnetic-Field-Insensitive Transition of $^{87}$Rb}
\author{R. Chicireanu}
\email[]{rchicireanu@gmail.com}
\affiliation{Joint Quantum Institute, National Institute of Standards and Technology and University of Maryland, Gaithersburg, Maryland 20899, USA}
\author{K. D. Nelson}
\affiliation{Joint Quantum Institute, National Institute of Standards and Technology and University of Maryland, Gaithersburg, Maryland 20899, USA}
\author{S. Olmschenk}
\affiliation{Joint Quantum Institute, National Institute of Standards and Technology and University of Maryland, Gaithersburg, Maryland 20899, USA}
\author{N. Lundblad}
\affiliation{Department of Physics and Astronomy, Bates College, Lewiston, Maine 04240, USA}
\author{A. Derevianko}
\affiliation{Department of Physics, University of Nevada, Reno, Nevada 89557, USA}
\author{J. V. Porto}
\affiliation{Joint Quantum Institute, National Institute of Standards and Technology and University of Maryland, Gaithersburg, Maryland 20899, USA}

\date{\today}
\begin{abstract}

We demonstrate near-complete cancellation of the differential light shift of a two-photon magnetic-field-insensitive microwave hyperfine (clock) transition in $^{87}$Rb atoms trapped in an optical lattice. Up to $95(2)\%$ of the differential light shift is cancelled while maintaining magnetic-field insensitivity. This technique should have applications in quantum information and frequency metrology.

\end{abstract}
\maketitle

Optical trapping of atoms is an indispensable tool for coherent quantum control of atomic spins. However, in many cases inhomogeneous differential light shifts (DLS) constitute a major limitation on spin-coherence times. Recent work~\cite{Derevianko01,Lundblad01,Vuletic01,Kuzmich01} has shown that in some cases the detrimental DLS in ground-state hyperfine levels of alkali atoms can be compensated. This approach has been used to increase spin coherence times~\cite{Vuletic01, Kuzmich01} and could impact quantum memories and computation,  as well as atom-based metrology.  In these demonstrations, the DLS cancellation has been obtained at the expense of Zeeman sensitivity to magnetic fields. In this paper, we experimentally explore the possibility of obtaining {\em simultaneous} DLS and magnetic field insensitivity. This idea has also been investigated theoretically in a recent proposal~\cite{Derevianko02}, which suggests that simultaneous cancellation of DLS and Zeeman shifts is possible for certain spin transitions.

The shift $\delta \nu$ of the transition frequency of an optically trapped atomic sample from its free-space, field-free value is determined by a combination of the electronic interaction with light and the electronic and nuclear Zeeman interaction with an external magnetic field, modified by the hyperfine interaction that couples the electronic and nuclear degrees of freedom. The resulting sensitivities $\partial \nu/\partial B$ and $\partial \nu/\partial I$, are functions of the magnetic field ${\bf B} =  B {\bf e}_{_B}$, and the intensity $I$ and polarization $\vec{ \epsilon}$ of the trapping light. For ground state hyperfine transitions of alkali metal atoms in the absence of light, $\partial \nu/\partial B$ is given by the Breit-Rabi formula~\cite{Breit-Rabi}. At low field, magnetic-field-insensitive   transitions ($\partial \nu/\partial B=0$) occur for the well known (single-photon) ``clock'' transitions $\ket{F, m_F=0} \leftrightarrow \ket{F', m_{F'}=0}$ at $B = 0$,  as well as for multiple-photon transitions $\ket{F, m_F} \leftrightarrow \ket{F', m_{F'}=-m_F}$ at certain non-zero ``magic'' $B = B_{\rm m}$ (see {\em e.g.}~\cite{Cornell01}). Here, $F$ denotes the atomic total angular momentum quantum number, and $m_F$ its projection on the quantization axis.

In the presence of light, the energy shifts $\Delta U$ giving rise to $\partial \nu/\partial I$ can be expressed as a sum of a scalar ($\Delta U^s$) and a vector ($\Delta U^v$) component~\cite{footnote1}. $\Delta U^s$ is rotationally invariant, depending only on $F$ and $F'$. $\Delta U^v$ depends on $m_F$ and $m_{F'}$ and to lowest order can be formally expressed as a differential Zeeman shift produced by a light-induced effective magnetic field ${\bf B}_{\rm eff}\equiv (E_{0}/2)^2 (\alpha_{F}^{v}/\mu_{F})(i \vec{\epsilon}^{*}\times \vec{\epsilon})$~\cite{Deutsch01}. Here $E_{0}$ and $\vec{\epsilon}$ are the amplitude and (complex) polarization unit vector of the electric field, such that $(i \vec{\epsilon}^{*}\times \vec{\epsilon})$ represents the direction and relative magnitude of the circular polarization, $\mu_{F}$ is the magnetic moment and $\alpha_{F}^{v}$ is the (possibly $B$-dependent) vector polarizability. In the limit where $B_{{\rm eff}} \ll B$, as is the case in our current experiment, only the component of ${\bf B}_{\rm eff}$ along ${\bf B}$ contributes to the shift, and

\begin{equation}\label{eq1}
\frac{\partial \nu}{\partial I}\propto (\alpha_{F'}^{s}-\alpha_{F}^{s})+\mathcal{A}\ (m_{F'} \alpha_{F'}^{v}- m_{F} \alpha_{F}^{v}),
\end{equation}
where $\alpha_{F}^{s}$ is the scalar component of the atomic polarizability, and $\mathcal{A}=(i \vec{\epsilon}^{*}\times \vec{\epsilon})\cdot {\bf e}_{_B}$ represents the degree of circularity of the polarization projected along ${\bf B}$.

For alkali ground state hyperfine transitions, there is no ``magic'' condition where the scalar differential shift ($\alpha_{F'}^{s}-\alpha_{F}^{s}$) vanishes, implying there is no way to cancel the shift with linearly polarized light~\cite{Rosenbusch01}, where $\mathcal{A}=0$. The technique used in~\cite{Lundblad01, Vuletic01, Kuzmich01} requires using the \emph{vector} light shift on a magnetic-field-sensitive transition to cancel the scalar DLS, requiring some component of circular polarization such that $\mathcal{A}\neq 0$.

It might appear impossible to use this technique to simultaneously cancel differential Zeeman and light shifts, as canceling the latter requires a non-zero vector DLS. Since the vector light shift acts as an effective magnetic field ${\bf B}_{\rm eff}$, we seemingly require a Zeeman sensitivity to ${\bf B}$.  The key insight is that the vector light shift (to lowest order) couples only to the {\em electronic} degree of freedom, whereas the Zeeman interaction includes a contribution from the nuclear magnetic moment. Therefore, the effective magnetic moments for the two interactions differ by roughly the nuclear magnetic moment, so at $B=B_{\rm m}$, where the differential Zeeman shift cancels, the differential vector light shift does {\em not} vanish. The vector DLS can still be used to counteract the scalar DLS.

We can derive approximate expressions for $\Delta U^s$ and $\Delta U^v$ to determine if cancellation is possible. For alkali atoms (with total angular momentum $F$ and $F'=F+1$ for the two hyperfine ground states), the Zeeman interaction $\widehat{H}_{\rm Z} = -(\mu_B  g_J {\bf \hat{J}} + \mu_N g_I  {\bf \hat{I}}) \cdot {\bf B}$, combined with the hyperfine interaction $\widehat{H}_{\rm HF} = A_{\rm HF} {\bf \hat{J} \cdot {\bf \hat{I}}}$, leads to field insensitive transitions at $B=B_{\rm m}$ for $m_{F+1}=-m_{F}$.  The light, however, interacts only with the electron angular momentum: $\widehat{H}_{\rm v}= -\mu_B g_J { \bf \hat{J}} \cdot {\bf B}_{\rm eff}$. This differs from the Zeeman Hamiltonian $\widehat{H}_{\rm Z}$ by the residual term $\mu_N g_I { \bf \hat{I}} \cdot {\bf B}_{\rm eff}$, where $g_I$ is the nuclear Land\'{e} factor. At $B=B_{\rm m}$, the vector DLS between states $|F^\prime,-m_F\rangle$ and $|F,m_F\rangle$ is $ \Delta U^v=-2 m_F \:\mu_N g_I B_{\rm eff}$~\cite{footnote2}. In terms of the total vector light shift $U^v$ of an individual hyperfine level, the differential vector light shift is $\Delta U^v \simeq -2 g_I \frac{\mu_N}{\mu_B} U^v$. The differential scalar shift relative to $U^v$ can be estimated as $\Delta U^{s} \simeq (3\Delta_{\rm HF}/ \Delta_{\rm F}) U^v$ \cite{Lundblad01}. For $^{87}$Rb, $ -2 g_I \frac{\mu_N}{\mu_B}=0.0020$ and $3 \Delta_{\rm HF}/ \Delta_{\rm F}=0.0029$, indicating $ \Delta U^v/  \Delta U^s$ is on the order of unity, and cancellation is possible. A more detailed calculation, including hyperfine corrections to the electronic wavefunctions, is needed to accurately determine the degree of cancellation. An accurate calculation for several alkali atoms shows that full cancellation for two-photon transitions is not possible, but in $^{87}$Rb the DLS sensitivity can still be decreased by a factor of 20~\cite{Derevianko02}.

\begin{figure}[t]
\begin{center}
\includegraphics[width=85mm]{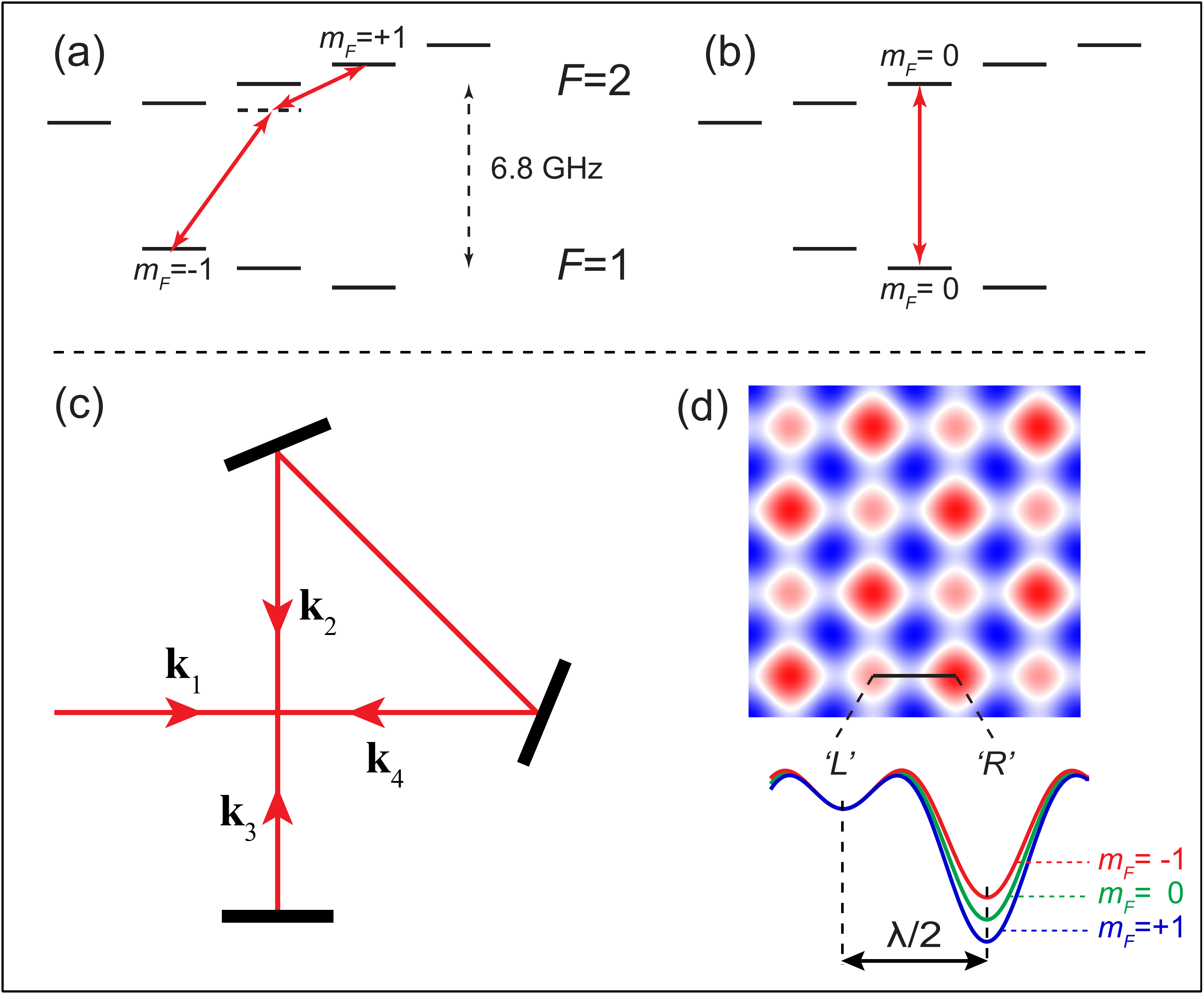}
\caption{Schematic of the $^{87}$Rb $\mu$w transitions studied here: (a) the two-photon transition $\ket{F=1,m_{F}=-1} \leftrightarrow \ket{F'=2,m_{F'}=+1}$, used to demonstrate the DLS reduction and (b) the $\ket{F=1,m_{F}=0} \leftrightarrow \ket{F'=2,m_{F'}=0}$ clock transition, used for comparison. In this experiment, atoms are trapped in an optical lattice formed at the intersection of four laser beams (${\bf k}_1$ to ${\bf k}_4$), obtained by a folded, retro-reflected single laser beam (c). The lattice intensity pattern used for DLS measurements is shown in (d), along with sections through the unit cell for three different $m_F$ states, and was experimentally optimized in order to maximize the amount of circular polarization on the right ($R$) sites. In this experiment atoms are trapped only in the $R$ sites.}
\label{fig:figure1}
\end{center}
\end{figure}

In the experiment described here, we study DLS of the $\ket{F=1,m_{F}=-1} \leftrightarrow \ket{F'=2,m_{F'}=+1}$ transition of $^{87}$Rb (see Fig.~\ref{fig:figure1}(a)) near $B_{\rm m}$=0.3228917(3)~mT~\cite{Cornell01}, by performing high-precision microwave ($\mu$w) spectroscopy of ultracold atoms trapped in an optical lattice. Measuring the transition frequency as a function of light intensity provides the sensitivity $\partial \nu / \partial I$. While it is generally relatively easy to measure the depth of an optical lattice (see~\cite{Ovchinnikov01}), it is more difficult to accurately measure the total intensity of the light at the position of the atoms. The main reason is the intensity imbalance between the counter-propagating lattice beams due to optical losses and beam area mismatch. To accurately quantify the expected reduction of the DLS, we therefore compare the two-photon $\mu$w transition with the single-photon $\mu$w transition between the $\ket{F=1,m_{F}=0}$ and $\ket{F=2,m_{F}=0}$ states. This transition, which is particularly important for its use in hyperfine atomic clocks, has a zero vector light shift at $B=0$.

Our atomic sample is produced by loading a $^{87}$Rb Bose-Einstein condensate (BEC) into a 3D optical lattice. We produce BECs containing typically $10^{5}$ atoms, spin-polarized in the $\ket{F=1,m_{F}=-1}$ state, with a duty cycle of $~ 60$~s. The final stage of the evaporative cooling is performed in a hybrid trap, created by a focused dipole laser beam  at 1550 nm and a quadrupole magnetic field slightly offset in the vertical direction $\hat{z}$ with respect to the center of the dipole beam, which provides longitudinal confinement and also compensates gravity (similar to \cite{YuJu01}). Subsequently, the atoms are adiabatically ($\sim$200~ms) loaded into the optical lattice, with a typical depth of about $30E_{\rm _R}$, undergoing the Mott insulator transition~\cite{Greiner01}. The recoil energy $E_{\rm _R}= \hbar^{2} k^{2}/2m_{{\rm _{Rb}}}$, where $k=2 \pi / \lambda$ is the lattice wave vector, and $m_{{\rm _{Rb}}}$ is the rubidium mass.

In order to avoid collisional shifts of the clock transition due to the state-dependent on-site interaction (which are of the same order of magnitude as the DLS), it is important to prepare only singly-occupied sites. To determine the atom number distribution in our lattice, we performed high-resolution two-photon spectroscopy of the clock transition \cite{Campbell01}, which allows us to differentiate between sites with different occupancies. To control the number of atoms in the BEC (before loading the lattice), we apply a non-adiabatic $\mu$w frequency sweep, removing a controlled fraction of the atoms by transferring them into the untrapped state $\ket{F=2,m_F=-2}$. For our trap parameters, we obtain single occupancy when the number of atoms in the BEC is less than $4\times 10^{4}$.

\begin{figure}[t]
\begin{center}
\includegraphics[width=75mm]{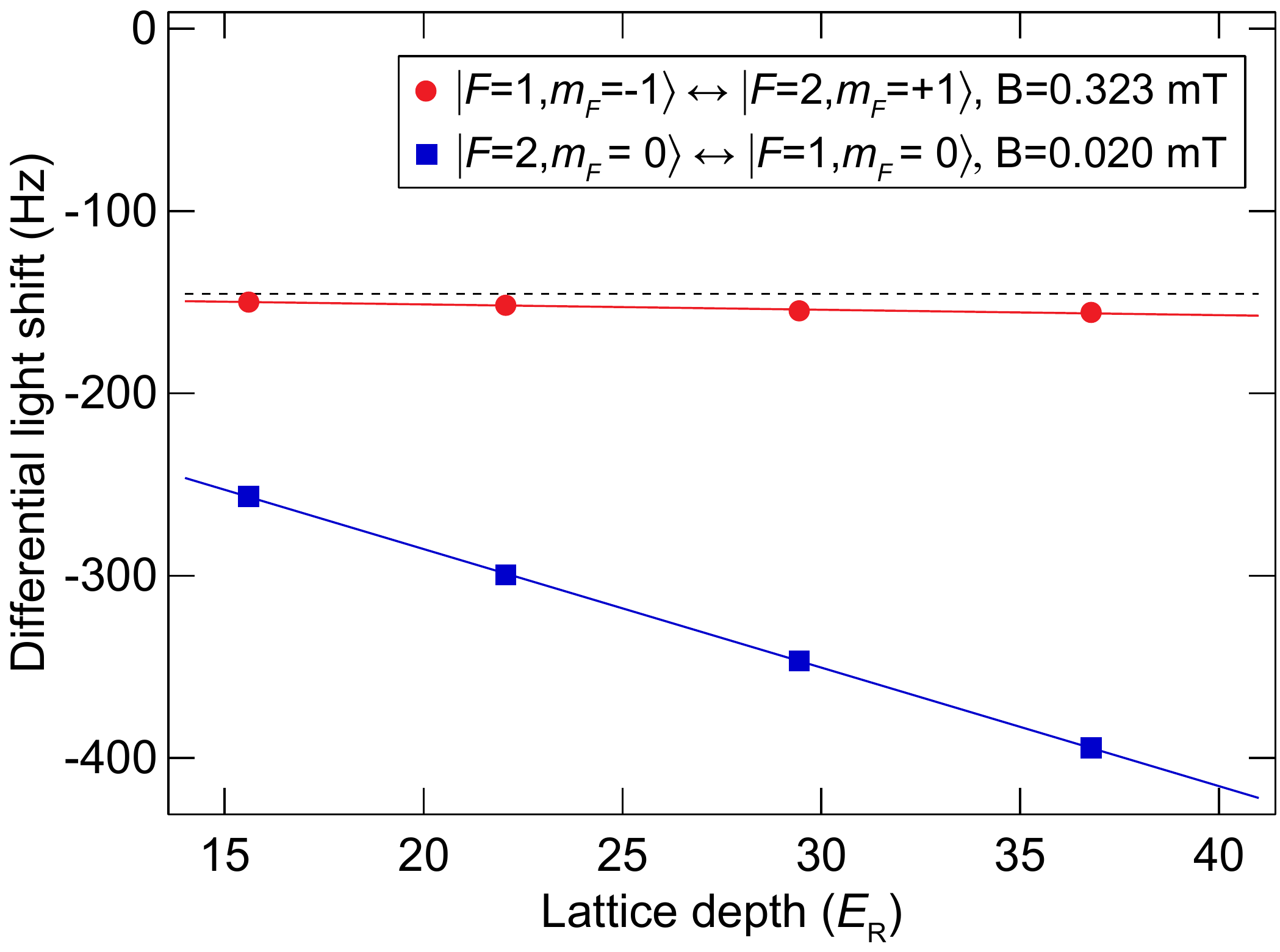}
\caption{DLS dependence on intensity, expressed as the lattice depth in the corresponding $\lambda$/2 configuration. We observe a significant reduction of the \emph{total} DLS of the two-photon clock transition (circles), compared to the single-photon clock transition (squares). The full lines represent linear fits to the data. The vertical lattice was kept at a constant depth of 30 $E_{{\rm R}}$, and the horizontal dashed line represents its (uncompensated) DLS, determined from extrapolating the fits to zero intensity. The statistical uncertainty of the data points is $\lesssim1$~Hz.}
\label{fig:figure2}
\end{center}
\end{figure}

The setup of our optical lattice has been described in detail elsewhere \cite{PLee01,Strabley01}. It consists of a 2D lattice in the horizontal ($\hat{x}-\hat{y}$) plane and an independent, linearly polarized vertical lattice (along $\hat{z}$), shifted in frequency by $\sim$180~MHz. The 2D lattice is obtained from a single, retro-reflected laser beam (Fig.~\ref{fig:figure1}(c)), and is adiabatically transformed during the experiment from a standard $\lambda/2$-period lattice with purely linear polarization (used during the loading stage) into the double-well configuration shown in Fig.~\ref{fig:figure1}(d) (used for spectroscopy), where the right ($R$) sites have an adjustable circular polarization component. ${\bf B}$ was aligned along the resulting  ${\bf B}_{\rm eff}$ (ideally in the $\hat{x}-\hat{y}$ plane), minimizing the DLS sensitivity on the $R$ sites. Based on the technique used in~\cite{PLee01}, we developed a procedure to trap atoms only in the $R$ sites: after loading the full lattice, we spectroscopically address the atoms on the left ($L$) sites and transfer them to the $F=2$ manifold, before expelling  them from the lattice with a resonant 20~$\mu$s light pulse, which does not affect the $F=1$ atoms in $R$.

To measure the transition frequencies, we use the detuned Ramsey method, consisting of two  $\pi/2$ pulses, separated by a variable hold time $\tau$, and with a typical detuning close to 1~kHz. We probe the two-photon transition using $\mu$w and rf fields, each detuned by 90~kHz from the intermediate $\ket{F=2,m_{F}=0}$ state (see Fig.~\ref{fig:figure1}(a)), resulting in a two-photon Rabi frequency of about 1~kHz. The single photon transition (Fig.~\ref{fig:figure1}(b)) is driven using a single $\mu$w field with a Rabi frequency of 9~kHz. After the Ramsey interrogation, state detection is performed by transferring the atoms between $\ket{F=1,m_{F}=-1}$ and  $\ket{F=2,m_{F}=0}$ with a $\mu$w $\pi$-pulse, switching off the lattice in $\sim$600~$\mu$s and absorption-imaging the cloud after 18~ms of time of flight and Stern-Gerlach separation.

\begin{figure}[t]
\begin{center}
\includegraphics[width=75mm]{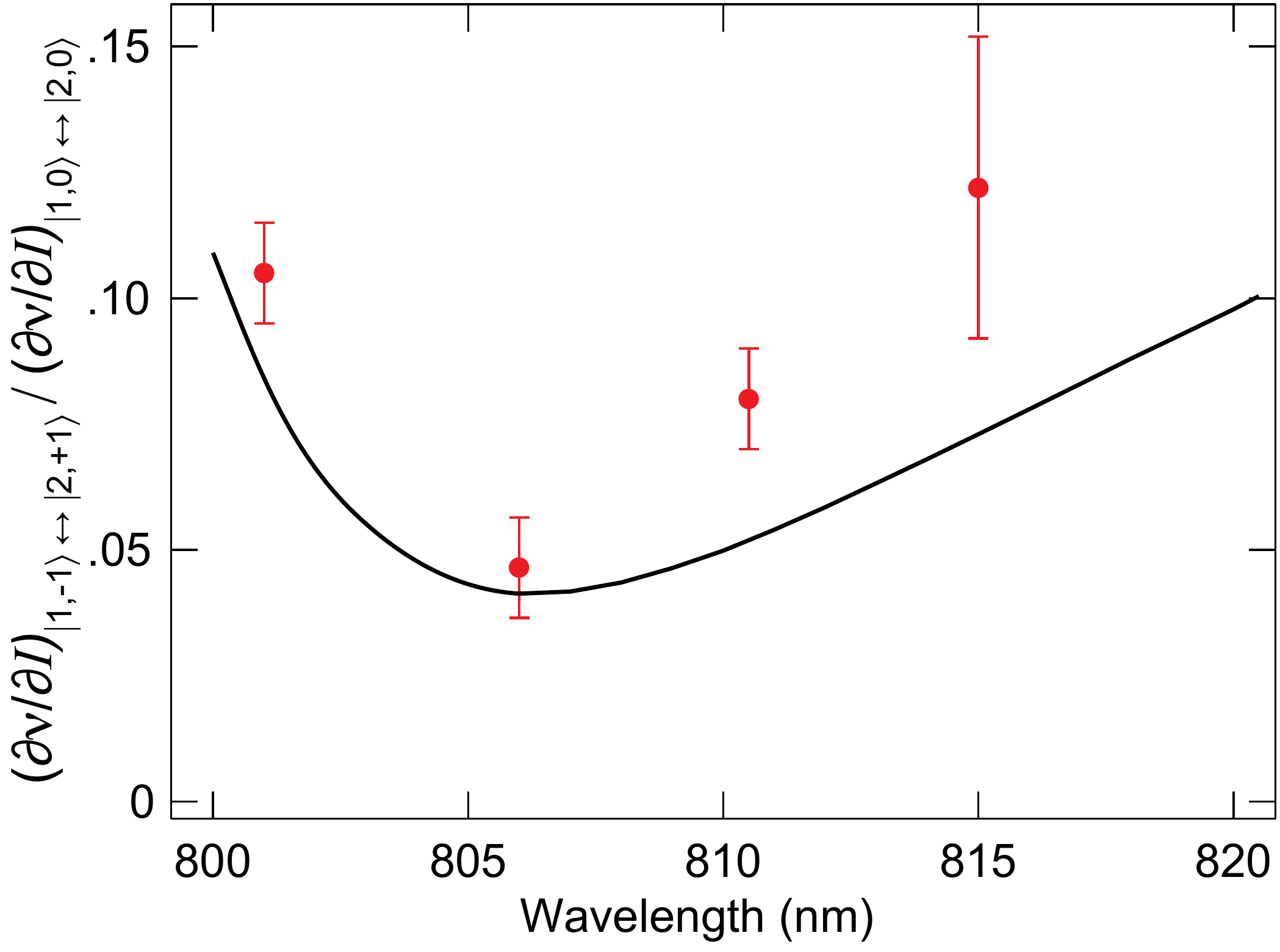}
\caption{Ratio between the DLS sensitivities ($\partial \nu / \partial I$) for the $\ket{F=1,m_{F}=-1} \leftrightarrow \ket{F'=2,m_{F'}=+1}$ and $\ket{F=1,m_{F}=0} \leftrightarrow \ket{F'=2,m_{F'}=0}$ transitions as a function of the lattice wavelength, showing a `nearly-magic' behavior, with a minimum value of $4.5\%$ at 806~nm. The magnetic field was kept at 0.323(3)~mT, near the magic value for the two-photon transition, whereas a small, 20~$\mu$T  bias was used for the single-photon transition. The full line represents a calculation using the method presented in~\cite{Derevianko02}.}
\label{fig:figure3}
\end{center}
\end{figure}

Fig.~\ref{fig:figure2} shows the DLS for both transitions as a function of the lattice intensity, expressed as the lattice depth in the corresponding $\lambda/2$ configuration. For circularly-polarized 2D lattice light at the $R$ sites we observe a significantly reduced sensitivity to the lattice intensity of the two-photon transition at $B_{\rm m}$, compared to the single-photon transition near $B=0$. The scalar DLS is the same in both cases, and the ratio of the two- and single-photon transition sensitivities quantifies the reduction of the DLS.

The circularity of the lattice light along ${\bf B}$ was optimized by minimizing the intensity dependence of the frequency. Based on an independent measurement of the losses of our lattice beams, we estimate that the light has a projected circularity $\mathcal{A} \simeq 0.99$. Moreover, by reversing the direction of $B$, we confirmed an increased DLS sensitivity of the two-photon transition, as in this case the scalar and vector components add together.

To preserve a well-defined quantization axis for the $\ket{F=2,m_{F}=0} \leftrightarrow \ket{F'=1,m_{F'}=0}$ transition, we maintained a small (20~$\mu$T) magnetic field at which the residual vector component of the DLS is calculated to be less than 0.6~Hz. In a lattice the dependence of $\delta \nu$ on intensity is not strictly linear, due to a zero-point energy offset~\cite{Taichenachev01}, but we estimate that for our parameter range the deviation from linearity is smaller than the measurement uncertainties. The zero-point energy does contribute to a slight shift in the extracted slopes, of $<4\%$ compared to a traveling wave of corresponding intensity, but this systematic shift does not contribute to the ratio of the single- and two-photon slopes.

\begin{figure}[t]
\begin{center}
\includegraphics[width=75mm]{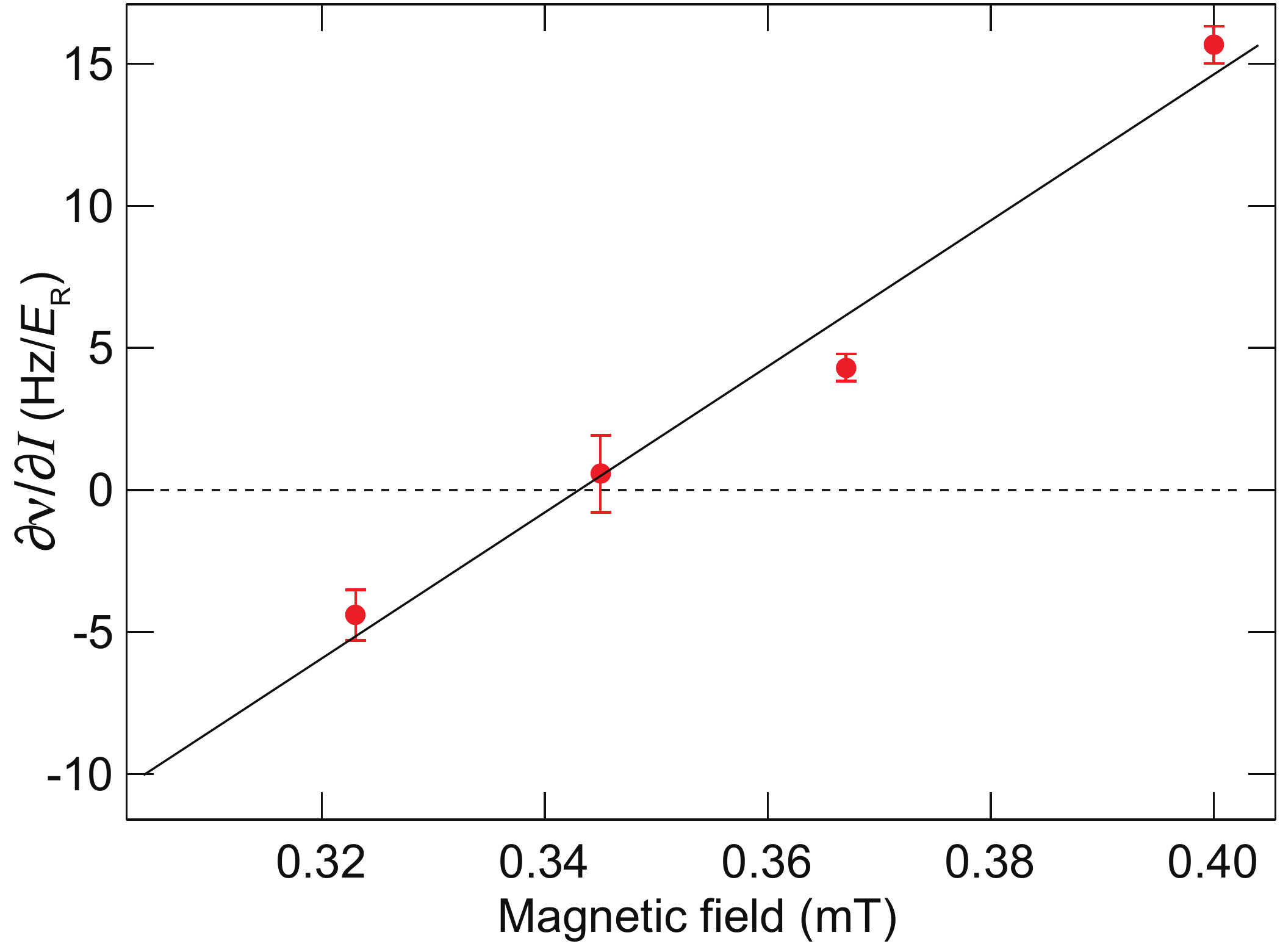}
\caption{Sensitivity of the transition frequency to the lattice laser intensity ($\partial \nu/\partial I$), as a function of the bias magnetic field, at $\lambda=806$~nm. At $B=0.343(3)$~mT (i.e. 20~$\mu$T away from the magic field) $\partial \nu/\partial I=0$, and the transition becomes insensitive, at first order, to fluctuations of the light intensity - with a residual field sensitivity $\partial \nu / \partial B$ of 1.7~Hz/$\mu$T.}
\label{fig:figure4}
\end{center}
\end{figure}

In Fig.~\ref{fig:figure3} we show the dependence of the $\partial \nu / \partial I$ ratio on the lattice wavelength between $802$~nm and $815$~nm. We observe a local minimum of the sensitivities ratio near 806~nm. This corresponds to a $\sim$$95(2)\%$ reduction of the DLS, in good agreement with our theoretical calculations (Fig.~\ref{fig:figure3}). The range of wavelengths accessible for this study was limited by two factors: at shorter wavelengths the measurement precision is limited by an enhanced rate of photon scattering from the lattice beams, while at longer wavelengths the DLS becomes comparable to our measurement uncertainties.

We also investigated the possibility of achieving $\partial \nu / \partial I=0$ by slightly shifting the magnetic field away from $B_{\rm m}$. By measuring the light shift sensitivity of the two-photon transition as a function of the magnetic field at $\lambda=806$~nm, we observe the linear dependence shown in Fig.~\ref{fig:figure4}. The DLS cancels completely at $0.343(3)$~mT ($\sim$20$~\mu$T away from $B_{\rm m}$), in agreement with theory~\cite{Derevianko02}. At this field, the residual magnetic field sensitivity is $\sim$1.7~Hz/$\mu$T, about  two orders of magnitude larger than what is typically used in atomic fountain clocks~\cite{Heavner01}.
 
We used the detuned Ramsey method, presented above, to compare the sensitivities of the two clock transitions to lattice inhomogeneities. We observed an increase in the coherence time between the two levels of the two-photon transition compared to the single-photon transition. At $\tau$=0 the Ramsey contrast is $98(2)\%$; at $\tau$=200~ms the single-photon contrast has decayed to zero, whereas the two-photon contrast is $10\%$. This value is likely limited by the inhomogeneous, uncompensated DLS of the vertical lattice, which is needed in our setup to support the atoms against gravity.

In summary, the experiments presented in this paper demonstrate a scheme to significantly reduce the light shift sensitivity of an atomic $\mu$w ground-state transition while retaining insensitivity to magnetic field fluctuations, using a subtle effect originating in the small difference between the \emph{total} and \emph{electronic} magnetic moments. While simultaneous full cancellation of both differential light and magnetic field shifts cannot be achieved for $^{87}$Rb, tuning experimental parameters between the differential light and Zeeman shift insensitive points may allow for minimizing the effect of the external field inhomogeneities and fluctuations on the coherence time of trapped atomic samples. The reduced sensitivity, demonstrated here in a 3D lattice, would be even more effective for applications using dipole traps and optical lattices in 1D and 2D geometries~\cite{Vuletic01,Kuzmich01,Polzik01}.

The experiments presented here confirm theoretical calculations of DLS~\cite{Derevianko02}. These calculations also predict perfect DLS cancellation at the magnetic-field-insensitive point for four-photon transitions in other alkali atoms ($^{85}$Rb and $^{133}$Cs). In these cases, cancellation may even be possible in a 3D lattice geometry by introducing circular polarization components along all coordinates and carefully chosing the external magnetic field orientation.

Ultimately, the reduced sensitivity to fluctuations demonstrated here, and its potential extension to other atoms outlined above, may have a range of applications including frequency metrology~\cite{Derevianko03}, coherent light storage~\cite{Kuzmich01}, and extending the coherence time of quantum memories for quantum information processing~\cite{Hammerer01}.

\subsection{ACKNOWLEDGMENTS}
We thank Saijun Wu for helpful discussions and Steven Maxwell, Ian Spielman and William D. Phillips for useful comments on this manuscript. This work is supported by the DARPA QUEST program. K.D.N. and S.O. acknowledge support from the National Research Council (NRC) Research Associateship program. Work of A.D. is supported in part by the NSF and by NASA under Grant/Cooperative Agreement No. NNX07AT65A issued by the Nevada NASA EPSCoR program.

\bibliographystyle{apsprl}

\begin{thebibliography}{29}
\bibitem{footnote1} The tensor DLS is negligible for the range of detunings used in our experiment, and is omitted from this paper.
\end{thebibliography}

\end{document}